\documentstyle[aps,prl,multicol,epsfig]{revtex}    

\newcommand{\beq}[1]{\begin{equation} \label{#1}}
\newcommand{\eeq}{\end{equation}}  
\newcommand{\beqa}[1]{\begin{eqnarray} \label{#1}}
\newcommand{\eeqa}{\end{eqnarray}}

\begin{document}

\title{Local size segregation in polydisperse hard sphere fluids}
\author{I. Pagonabarraga, M.E. Cates and G.J. Ackland}
\address{Department of Physics and Astronomy, JCMB The King's Buildings,\\
 Mayfield Road, EH9 3JZ Edinburgh, Scotland}
\maketitle
\date{}

\begin{abstract}
The structure of polydisperse hard sphere fluids, in the presence of a wall, is 
studied by the Rosenfeld density functional theory. Within this approach, the local 
excess free energy depends on only four combinations of the full set of density fields. 
The case of continuous polydispersity thereby becomes tractable. We predict, generically, 
an oscillatory size segregation close to the wall, and connect this, by a perturbation 
theory for narrow distributions, with the reversible work for changing the size of one 
particle in a monodisperse reference fluid. 
\end{abstract}

\begin{multicols}{2}
\narrowtext

Understanding the behavior of polydisperse systems is relevant to many materials of practical interest. 
In particular, colloidal and/or polymeric fluids generally contain particles which have, in effect, a 
continuous distribution of sizes (and/or other parameters such as charge and chemical composition). 
This affects their performance in applications ranging from foodstuffs to polymer 
processing\cite{general}. More fundamentally, colloidal systems also provide the closest experimental 
approach to the `theorists ideal fluid', namely that of perfect hard spheres\cite{puseyrev}. The fact 
that all colloids are in practice polydisperse (at least slightly) must then be taken into account in 
comparing theory with experiment. Only recently has experimental work started to clarify in a systematic
 way the generic consequences of polydispersity, such as the partitioning of sizes between coexisting 
phases\cite{dave}.

Despite the continuous interest they have raised, the theoretical understanding of polydisperse fluids 
remains far from complete, especially for inhomogeneous cases. More is known about partial structure 
factors in single-phase fluids \cite{Lado,klein} and liquid-liquid phase 
equilibrium\cite{salacuse,gualteri} than about crystalline phases\cite{terminal} or interfacial 
properties, for example. And, where such inhomogeneous situations have been studied \cite{mcrae} it has 
often proved necessary to assume that only the mean density, and not the size distribution, can vary in 
space \cite{kofke}. This ignores size segregation effects, which (globally) influence the phase 
diagram\cite{gualteri,dave}. A similar tendency to {\em local} segregation is implicit in treatments 
of binary and ternary hard sphere mixtures\cite{Dijkstra1,binary,preprint} and in polydisperse 
equilibrium structure factors in the homogeneous state\cite{Lado}.

In what follows, we treat continuous polydispersity within a density functional theory (DFT)
 that properly allows for local size segregation. Our work is based on a choice of density functional 
(that of Rosenfeld\cite{Rosenfeld}) that has previously been used to study finite mixtures of hard 
spheres. By exploiting the fact that its excess free energy density depends on only a small number of 
linear combinations of the particle densities (four `moment densities'), we are able to 
address the case of continuous polydispersity, where the underlying densities are infinite in number. 
This allows us to study, e.g., the effects of varying the {\em shape} of a smooth size 
distribution.  Moreover, by a perturbative analysis of the same functional, we can distinguish certain 
generic features that do not depend on the shape of the parent, if it remains narrow. These two aspects 
of our work build on recent nonperturbative\cite{sollich} and perturbative\cite{dave} progress 
in understanding polydisperse phase equilibria, extending both approaches to the case of inhomogeneous 
fluids for the first time.

The state of a polydisperse fluid is specified by the local number density of each species, 
$\rho(\sigma,{\bf r})$, with $\sigma$ the particle diameter (say). The spatial average of this quantity 
must recover the (known) global size distribution, $\rho(\sigma)$, which we call the `parent'. 
According to density functional theory, the grand potential is some (unknown) functional of 
$\rho(\sigma,{\bf r})$. Given an approximation to this functional, the key problem of polydispersity is 
the need to find, by its minimization, an infinite set of densities at each point in space. A naive 
discretization into (say) $N=50$ species does not make this much easier: the minimization problem 
remains of very high dimension (we have $N$ functions of $d$ spatial coordinates, with $d$ the effective
 space dimension; $d =1$ for a fluid near a flat wall).

The problem of free energy minimization in a very large space arises already in the calculation of 
phase diagrams, where it has been found that, for many purposes, the problem can be exactly projected 
onto a subspace involving a few linear combinations (or `moments') of $\rho(\sigma)$ \cite{sollich}. 
The approach requires that the {\em excess} free energy can be expressed in terms of the chosen 
moments; this is true for many approximate theories including polymer mean-field theories and liquid 
state models such as the Percus-Yevick equation of state (PY)\cite{B_Evans,PBWepl}.

We now observe that essentially the same simplification is possible with certain (approximate, but 
well-established) density functionals for inhomogeneous liquids \cite{B_Evans}. We choose for 
definiteness that of Rosenfeld\cite{Rosenfeld} for hard spheres, as subsequently recast by Kierklik 
and Rosinberg\cite{Kierlik}, whose non-ideal part is a functional of four `moment-densities' defined, 
from the underlying density profile $\rho(\sigma,{\bf r})$, as follows (with $\alpha = 0,1,2,3$):
\begin{equation}
m_{\alpha}({\bf r}) =\int d\sigma d{\bf r}'\rho(\sigma,{\bf r}') 
\omega_{\alpha}(\sigma,|{\bf r}'-{\bf r}|) \label{moments} 
\end{equation}
Here $\omega_{\alpha}(\sigma,|{\bf r}'-{\bf r}|)$ are four weight functions, 
selected\cite{Rosenfeld,Kierlik} such that PY is recovered for a homogeneous mixture. Note that only 
four moment densities are needed, irrespective of the number of components of the mixture; this follows 
from PY itself, whose nonideal part ${\cal F}^{ex}(m_\alpha)$ involves only the four moments 
$m_\alpha =\int d\sigma \rho(\sigma) \sigma^\alpha$. However, the weight functions $\omega_\alpha$ are 
nonlocal: our `moment densities' $m_\alpha({\bf r})$, though intimately related to the four PY moments, 
are not merely local values of them.
 
Within this description, the grand potential becomes
\begin{eqnarray}
\Omega &=& \int d{\bf r}d\sigma \left\{\rho (\sigma,
{\bf r}) \left[\ln (\Lambda^3(\sigma) \rho (\sigma,{\bf r}))-1\right] + \right.\nonumber\\
&&\left. {\cal F}^{ex}\left(m_{\alpha}({\bf r})\right)
 +(V(\sigma,{\bf r})-\mu(\sigma)) \rho (\sigma,{\bf r}) \right\}
\label{functional}
\end{eqnarray}
where $\mu(\sigma)$ is the chemical potential of species $\sigma$ in the 
bulk, $\Lambda(\sigma)$ is a thermal wavelength\cite{salacuse}, $V(\sigma,{\bf r})$ is the external potential acting on species 
$\sigma$, and ${\cal F}^{ex}(m_{\alpha}) = -m_0\ln(1-m_3) + m_1m_2/(1-m_3) + m_2^3/(24\pi(1-m_3)^2)$ 
gives the usual PY result\cite{B_Evans}. Minimization of eq.(\ref{functional}) leads to eq.(\ref{moments}) with
\begin{equation}
\rho(\sigma,{\bf r}) = R(\sigma)\exp\left\{-\beta V(\sigma,{\bf r}) +
\tilde c(\sigma,{\bf r})-\tilde c(\sigma,\infty)\right\}
\label{eq_mom}
\end{equation}
\begin{equation}
\tilde c(\sigma,{\bf r})=-\sum_{\beta}\int d{\bf r}' 
\frac{\partial {\cal F}^{ex}}{\partial m_{\beta}}({\bf r}') \omega_{\beta}(\sigma,
 |{\bf r}'-{\bf r}|)\label{eq_c}
\end{equation}
Here $\tilde c(\sigma,\infty)$ is the value of the excess chemical 
potential, and $R(\sigma)$ the density distribution, in the bulk.

For a localized potential $V$ (such as a hard wall), and for any other case where the densities differ 
from their bulk values only in a finite neighborhood, $R(\sigma)$ coincides with the parent 
$\rho(\sigma)$ in the thermodynamic limit\cite{footnote}, and is known in advance. 
Eqs.(\ref{moments},\ref{eq_mom},\ref{eq_c}) are then {\em closed} in the low-dimensional function space 
spanned by $\left\{m_\alpha({\bf r})\right\}$. Their numerical evaluation delivers 
(via eq.(\ref{eq_mom})) the full density profile, but nonetheless proceeds as if there were only four 
species present. Note that the four moment-density profiles depend (through eqs.(\ref{moments},
\ref{eq_mom})) on $R(\sigma)$, so that the final results, even for the $m_\alpha({\bf r})$ themselves, 
still depend on {\em all} moments of the parent distribution. 

Of course, not all density functionals in common use are of the required `moment density' 
form\cite{B_Evans}. But eq.(\ref{functional}) performs as well as most other functionals proposed in 
the literature\cite{preprint}; for example, depletion forces in binary fluids are well recovered by 
it\cite{Dijkstra}. For numerical use below we retain the original weight functions \cite{Rosenfeld}, 
although some recent modifications are known to give a better description of solid phases \cite{Lowen}; 
these modifications would not alter the conceptual structure of our analysis.

We now focus on the effect of polydispersity in a fluid of hard spheres near a flat hard wall. We fix 
the parent size distribution $R(\sigma)= \rho(\sigma,\infty)$, and thereby the chemical potentials of all 
species in the bulk, and analyse the structure as a function of distance $z$ from the wall.
We have considered fluids with a Gaussian, an exponential, and a uniform (top hat) distribution of 
sizes; each is characterized by its mean (which we set as $\bar \sigma = 1$ without loss of generality) 
and its standard deviation (which is then called the polydispersity, $p$, and is expressed in \%). We disregard any transverse ordering, which is reasonable for a fluid, except possibly close to the fluid/solid transition.

\begin{center}
\begin{figure}[tbh]
\vspace{-0.5cm}
\epsfig{figure=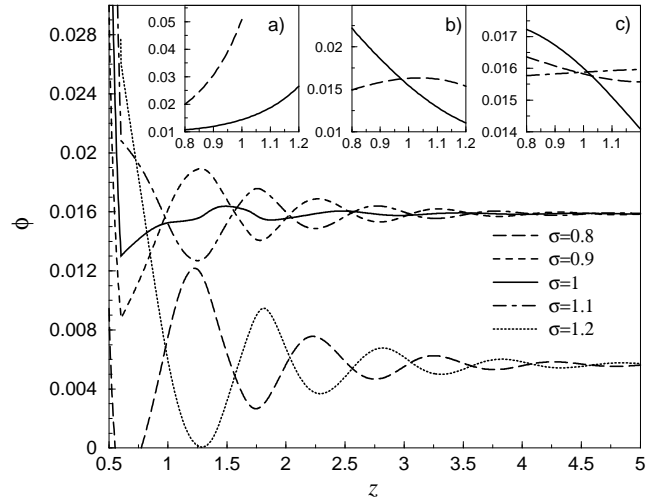,width=8.5cm,angle=0}
\caption{ Profiles for the local concentration of several species for a flat parent with 
$p=11.5\%$, at mean volume fraction $\eta=0.4$. The mean diameter is taken as 
the length unit. The curves for $\sigma=0.8$ and $\sigma=1.2$ have been vertically 
displaced to aid comparison. Insets: Size distribution at 
different $z$. a) Solid line, $z=0.5$; dashed line, $z=0.8$. b)  
Solid line, $z=1.2$; dashed line, $z=1.6$. c) Solid line, 
$z=2.4$; dashed line, $z=3.2$; dotted-dashed line, $z=4.7$.}
\label{fig:phi}
\end{figure}
\end{center}

For all the parents considered, we find significant local size segregation, with a strong, local cross 
correlation between particles of different sizes. This effect can be observed in fig.\ref{fig:phi}, 
where we display the {\em local relative concentration} of various species, 
$\phi(\sigma,z) = \rho(\sigma,z)/\int d\sigma \rho(\sigma,z)$, for the case of a top-hat parent with 
$p=11.5\%$. Clearly visible are strong, anticorrelated oscillations in the relative amounts 
of large and small particles, whereas the relative concentration of particles close to the mean size 
($\sigma = 1$) is much more nearly constant. Our definition of $\phi(\sigma)$ factors out the 
primary oscillations in the mean density (depicted in fig.\ref{fig:scalephi} below), but we find that 
the size distribution (fig.\ref{fig:phi}, insets) oscillates roughly in quadrature to the mean density. 
In this manner, without greatly altering the overall density profile, smaller particles are accommodated
 on the inner (near-wall) side of each successive density peak and larger particles on the outer side. 
Thus the density of each species can, within the first few layers at least, oscillate with a spatial 
period close to its own diameter. (The same could not continue indefinitely: the nature of any size 
segregation in ordered phases thus remains an important, open issue. Moreover, the 
asymptotic decay of the density profile for any mixture is given by a single wavelength\cite{evans-leote}).
 As the polydispersity of the parent increases, the trend we have outlined persists, 
although it blurs, and the uniform relative concentration of the mean species is lost. The behavior 
is generic for all the different size distributions tried, and can also be seen in ternary
 (or even binary) mixtures\cite{preprint}. 

We can gain further insight into this behavior by  a perturbative analysis for a narrow parent, 
analogous to that used previously for phase coexistence \cite{gualteri,dave}. We 
express the particle diameters as $\sigma=\bar{\sigma}(1+\epsilon)$; we then expand in small $\epsilon$ around a monodisperse density profile for that 
species. The perturbative result is (to order $\epsilon$)\cite{preprint}
\begin{eqnarray}
\rho (\sigma,z) &=& \frac{R(\sigma)}{\rho^{(m)}(\infty)} 
\rho^{(m)}(z) e^{-\beta (V(\sigma,z)-
V(\bar{\sigma},z))}\nonumber\\
&& \times \left(1+\epsilon \left[\tilde{c}'(z)-\tilde{c}'(\infty)\right]\right) 
\label{rho}
\end{eqnarray}
Here $\rho^{(m)}(z)$ is the density profile for the monodisperse system, and $\tilde{c}'(z)$ is the 
reversible work coefficient for slightly changing the size of one particle, $z$ from the wall, within 
an otherwise monodisperse system. This obeys: 
\begin{equation}
\tilde{c}'({\bf r}) = \frac{\partial}{\partial\epsilon}\tilde{c}(\sigma,{\bf r})|_{\epsilon=0,p\rightarrow 0}
\label{muex}
\end{equation}
which can be easily computed from eq.(\ref{eq_c}). We have assumed that $\epsilon \ll 1$ for 
all members of the parent, but not that $\epsilon\ll p$; this is why there is no expansion made of the 
factor $R(\sigma)$ in eq.\ref{rho}. Likewise for a hard wall it is inappropriate to expand 
$V(\sigma,z)$.
\begin{center}
\begin{figure}[tbh]
\epsfig{figure=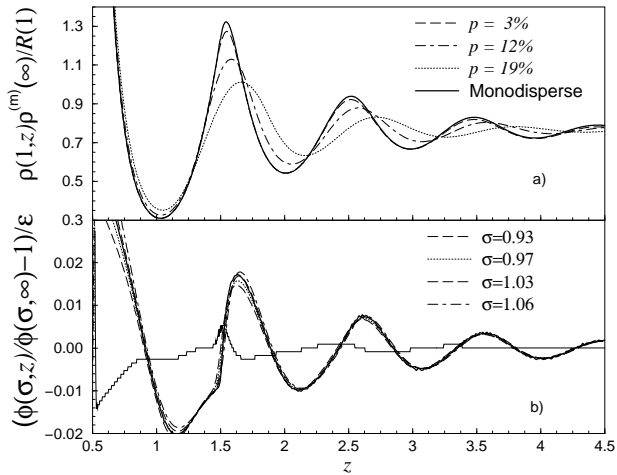,width=8cm,angle=0}
\caption{ a) Density profiles for the mean species with a flat parent, setting $\bar{\sigma}$=1,
 compared with the profile of a monodisperse fluid at the same mean volume 
fraction, $\eta=0.4$. b) Reduced local concentration profiles, for a Gaussian parent ($p$ = 2.4\%) at 
 $\eta=0.45$, compared with $\tilde c'(z) -\tilde c'(\infty)$ from the 
monodisperse fluid at the same $\eta$, thick continuous line. 
The thin continuous line refers to the local 
concentration of the mean species. This curve is not reduced, but has been shifted vertically so 
that it tends to zero at large $z$. 
\label{fig:scalephi}}
\end{figure}
\end{center}

These perturbative calculations predict that the profile of the `mean species' is, to order $\epsilon$, 
identical to a monodisperse fluid at the same overall density. This agrees with fig.\ref{fig:scalephi}a,
 where we compare the mean-species profiles for flat parents of varying polydispersity. 
Curves for other parents (not shown), but with matched polydispersity, are barely distinguishable on 
such a plot, even when this is over 10\%. More generally, eq.(\ref{rho}) shows that all the order 
$\epsilon$ deviations from monodisperse behavior, for parents of different shapes, should fall on a 
common curve if the densities $\rho(\sigma,z)$ are first normalized by $R(\sigma)$.
We have verified this by comparing $\rho(\sigma,z)/R(\sigma)$ for small fixed $\epsilon$, among 
parents of different shapes\cite{preprint}. 

An illuminating application of the perturbation theory is to the local concentration profiles shown in 
fig.\ref{fig:phi}. According to eq.(\ref{rho}), in any region where the external potential does not depend on the radius of the particles,
\begin{equation}
\phi(\sigma,z) = \phi(\sigma,\infty)\left[1+\epsilon \left(\tilde c'(z)-\tilde 
c'(\infty) \right) \right]
\end{equation}
which shows, to order $\epsilon$, that the local concentration of the mean species is strictly constant, 
and that by an appropriate scaling, the concentration profiles for different particle sizes collapse
onto a single curve. These predictions are confirmed from the full (nonperturbative) 
solution for a narrow Gaussian parent in fig.\ref{fig:scalephi}b. The size oscillations are, as 
predicted, directly linked to the reversible work term, $\tilde c'(z)$, for changing the size
of one particle. Although the data
 collapse is mostly excellent, the constancy of the mean species concentration is imperfect close to 
the wall even for 
narrow (2.4\%) polydispersity. This may be because the perturbative expansion itself fails very 
close to a hard wall, where the potential forces the concentration of certain species to be zero. 
(This is visible in the first  inset of fig.\ref{fig:phi}.) The same problem should not arise for a 
smoothly varying potential.

\begin{center}
\begin{figure}[tbh]
\vspace{-0.5cm}
\epsfig{figure=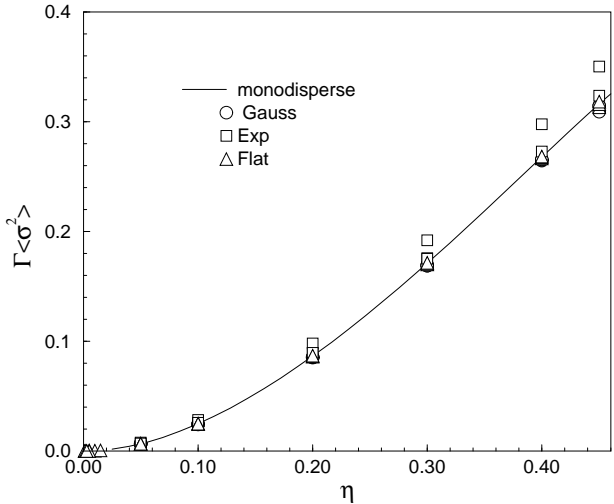,width=8cm,angle=0}
\caption{Adsorption for different parents. Gaussian: 2.4\%, 7\%, 12\%; exponential: 9\%, 23\% (cutoff at a minimum $\sigma$ of 0.91 and 0.77 respectively), flat: 4\%, 12\%, 19\%.}
\label{fig:ads1}
\end{figure}
\end{center}

We now turn to thermodynamic properties of our system. We have 
integrated the density and the energy profiles to find the adsorption $\Gamma$ and the surface 
tension\cite{B_Evans}. In fig.\ref{fig:ads1} we show the $\Gamma$ as a function of the volume fraction, 
for different parents and polydispersities. It appears that, until the parent becomes relatively wide, 
the values of the measured adsorptions do not differ significantly from the monodisperse case. 
Indeed, our perturbative calculations show the deviation, and that of the surface 
tension\cite{preprint}, to be of order $p^2$. But in fact the deviations are  {\em only} small if, as 
shown, one uses well-chosen moments to scale the plot (effectively, $\Gamma m_2/m_0$ vs $m_3$). These 
scalings are suggested by (e.g.) scaled particle theories for the adsorption\cite{spt}; with different 
choices, there are deviations of up to 30\% in the same data. 

In this Letter we have considered the effect of polydispersity on the structure of an inhomogeneous 
hard sphere fluid. We have shown that the use of moment densities makes DFT a practical tool to study 
inhomogeneous fluids of continuous polydispersity, in which local size segregation can play a major role. This was 
illustrated by a study of a polydisperse hard sphere fluid near a wall, which clearly shows such 
effects; spheres segregate so as to have density maxima with a period close to their own diameter. 
The generic character of this segregation was confirmed by a perturbative analysis.
Some of our order $\epsilon$ size segregation results have analogues in bulk phase fractionation of 
polydisperse fluids. We showed that the density of the mean species is unperturbed by weak 
polydispersity; we can also obtain a general relation that connects the ratios of differences in the 
local moments (evaluated from $\rho(\sigma,{\bf r})$ at two different spatial points) to ratios of 
higher moments evaluated for the parent \cite{preprint}. Such findings generalize
similar ${\cal O}(\epsilon)$ results for phase partitioning \cite{dave}; for these purposes, it 
is as if, in DFT, each point in space counts as a different `phase'. 

We note finally that some of our perturbative results (including those just described) are 
surprisingly general. For example, eq.(\ref{rho}), which connects the species densities to the local 
reversible work of enlarging a particle in the monodisperse limit, makes no assumption about the 
choice of density functional. Hence it holds for the {\em true} functional, whatever it may 
be \cite{diff}. Thus the choice of physical system (hard spheres), and of approximate functional 
(eq.\ref{functional}) enters only through the particular form of $\tilde c'({\bf r})$ (eq.\ref{muex}).
 Eq.(\ref{rho}) is thus an exact result for any slightly polydisperse isotropic fluid; moreover the 
polydisperse feature ($\sigma$) need not even be size, but could be charge, or any other scalar 
quantity.
\acknowledgments
We are grateful to P. Bartlett, R.M.L. Evans,
 P. Sollich and  P.B. Warren for their useful comments. Work funded by EPSRC GR M/29696.

\end{multicols}

\begin{thebibliography}{999}
\bibitem{footnote} The projection method of \cite{sollich} likewise becomes exact (for systems 
whose excess free energy has the required form) whenever only a small subvolume deviates (perhaps 
strongly) from the parent.
\bibitem{diff} This assumes the relevant derivatives of the exact density functional are finite, 
even for hard spheres.
\end{thebibliography}
\end{document}